\documentclass[aip,amsmath,amssymb,reprint]{revtex4-1}
\usepackage[dvipdfmx]{graphicx,hyperref}
\usepackage{dcolumn}
\usepackage{bm}
\usepackage[utf8]{inputenc}
\usepackage[T1]{fontenc}
\usepackage{mathptmx}
\usepackage{color}
\usepackage{graphicx}
\usepackage{dcolumn}
\usepackage{bm}
\usepackage{amsmath}
\usepackage{amsfonts}
\usepackage{amssymb}
\usepackage{comment}
\usepackage{siunitx}

\usepackage{etoolbox}

\makeatletter
\def\@email#1#2{%
 \endgroup
 \patchcmd{\titleblock@produce}
  {\frontmatter@RRAPformat}
  {\frontmatter@RRAPformat{\produce@RRAP{*#1\href{mailto:#2}{#2}}}\frontmatter@RRAPformat}
  {}{}
}%
\makeatother

\begin{document}

\preprint{AIP/123-QED}

\title{Pattern transition of flow dynamics in a highly water-absorbent granular bed}

\author{Kojiro Otoguro}
 \affiliation{Department of Applied Physics, Faculty of Science Division I, Tokyo University of Science, 6-3-1 Nijuku, Katsushika-ku, Tokyo 125-8585, Japan}%
 
\author{Kiwamu Yoshii}
 \affiliation{Department of Mechanical Science and Bioengineering, Graduate School of Engineering Science, Osaka University, 1--3 Machikaneyama, Toyonaka, Osaka 560--8531, Japan}% 
 \affiliation{Department of Physics, Nagoya University, Furo-cho, Chikusa-ku, Nagoya 464-8602 Japan}%

\author{Yutaka Sumino}
 \email[Authors to whom correspondence should be addressed: ]{ysumino@rs.tus.ac.jp}%
 \affiliation{Department of Applied Physics, Faculty of Science Division I, Tokyo University of Science, 6-3-1 Nijuku, Katsushika-ku, Tokyo 125-8585, Japan}%
 \affiliation{WaTUS and DCIS, Research Institute for Science \& Technology, Tokyo University of Science, 6-3-1 Nijuku, Katsushika-ku, Tokyo 125-8585, Japan}%

\date{\today}

\begin{abstract}
An aqueous sodium chloride solution was injected at a controlled rate into a granular bed in a quasi-two-dimensional cell. The granular bed was made of dried, highly water-absorbent gel particles whose swelling rate was controlled by the salinity of the injected fluid. At a high salinity level (low swelling rate), high injection rate, and short timescale, the injected fluid percolated between the gel particles in an isotropic manner. Meanwhile, at a low salinity level (high swelling rate), low injection rate, and long timescale, the gel particles clogged the flow path, resulting in anisotropic branch-like structures of the injected fluid front. The transition of the injection pattern could be understood based on the ratio of the characteristic timescales of swelling and injection. Moreover, the clogged pattern showed an oscillatory pressure drop whose amplitude was increased with higher salinity. Such an oscillatory behavior observed in an injection process in a swelling gel particle may be relevant in geological situation; i.e., such as fluid migration underground.
\end{abstract}

\maketitle

\section{Introduction}
When a highly viscous fluid filled in a quasi-two-dimensional cell, i.e., a Hele-Shaw cell, is replaced by a less viscous fluid, a characteristic fingering pattern is observed. This fingering pattern is referred to as viscous fingering and is caused by the Saffman-Taylor instability~\cite{saffman1958penetration}. The key concept of this pattern formation phenomenon is Darcy's law, which states that the flow is proportional to the pressure gradient~\cite{saffer2011hydrogeology}.
Recent studies on fingering patterns with a quasi-two-dimensional cell have been conducted under unique conditions such as granular beds~\cite{huang2012granular}, viscosity changes owing to chemical reactions~\cite{eckert2004chemical, haudin2014spiral,wagatsuma2017pattern,Tanaka2023,wang2021effect}, the effect of buoyancy~\cite{Haudin2014a}, variations in interfacial energy between the substrate and inner and outer fluids~\cite{Yoshii2024}, and miscibility of the fluid~\cite{mishra2009influence}.

One relevant application of this phenomenon is the motion of underground fluids.
The process of injecting fluids into the ground is used for chemical grouting~\cite{Karol2003}, oil recovery~\cite{Muggeridge2014}, and dealing with wasted liquefied carbon dioxide~\cite{Berg2012}.
These processes can result in severe injection-induced earthquakes~\cite{ellsworth2013injection}. The timescales and flow patterns beneath the ground provide relevant information for estimating the distance that the injected fluids can migrate and for predicting the nucleation of injection-induced earthquakes.
Thus, it is desirable to elucidate the flow patterns wherein the fluidity changes in a laboratory.

In a realistic underground situation, permeability can be changed by chemical reactions such as silica precipitation~\cite{Audet2014-po}. In many previous studies, fluid injection with chemical reaction-induced precipitate formation has been conducted~\cite{Podgorski2007, Haudin2014,haudin2014spiral,wagatsuma2017pattern,Tanaka2023, Nagatsu2008}.
The experiments were conducted under fluid-filled conditions, although these conditions differ from those of bedrocks with complex porous structures. Furthermore, estimating and controlling the dynamics of the precipitation reactions is not straightforward. Thus, this study focused on an experimental system with granular media composed of ionic gel particles with a high water absorption capacity.

Ionic gel particles can be swollen up to 10 times their length, which implies a 10$^3$-volume increase. Note that the volume change is comparable to that of the liquid-gas phase transition. The high absorption capacity of ionic gels is derived from the electrostatic repulsion between the charges on the polymer chains and the osmotic pressure caused by counterions on the polymer chains~\cite{tanaka1980phase}. Therefore, the absorbent capacity and swelling rate can be systematically altered by adding salt to water. Previous studies have used such highly swollen gel particles to create mechanical stress~\cite{Louf2021} in a granular bed, which may lead to the blocking of injected fluids. However, the size of the gel particle was on the \SI{}{\milli \meter} scale, and the swelling that occurred in the order of 100 h was too slow to conduct an injection experiment on a realistic timescale. In this study, we used a few \SI{100}{\micro \meter} particles instead of \SI{}{\milli \meter} particles to achieve a faster swelling rate at a speed comparable to that of the injection.

Here, we prepared Hele-Shaw cells filled with ionic gel particles and injected them with an aqueous salt solution. The adsorbent capacity and swelling rate of the gel particles were controlled by varying the salt concentration. Using salt concentration and injection rate as control parameters, we observed the injection patterns and found isotropic percolation and anisotropic clogged patterns. We examined the injection pressure and confirmed the appearance of oscillatory behavior whose amplitude increases for high salinity of injected fluid. We discuss possible cause of this pressure oscillation in terms of increased clogging region for high salinity.

%%%%%%%%%%%%%%%%%%%%%%%%%%%%%%%%%%%%%%%%%%%%%%%%%%%%%
\begin{figure}[ht]
\centering
\includegraphics[width=0.48\textwidth]{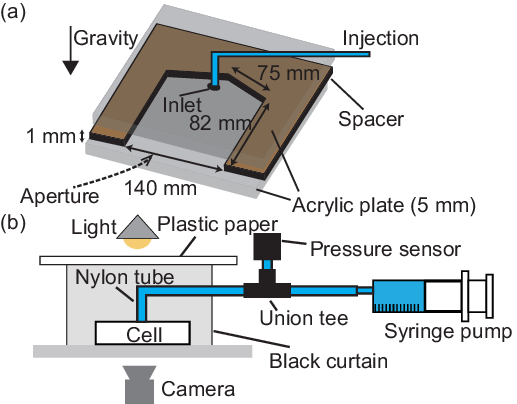}
\caption{\label{experimentalsystem} (a) Setup of Hele-Shaw cell with gap width set as $1$ mm. The cell was filled with gel particles and was placed horizontally. (b) Schematic of the assembled experimental setup. A fluid was injected with a syringe pump with a constant injection rate.}
\end{figure}
%%%%%%%%%%%%%%%%%%%%%%%%%%%%%%%%%%%%%%%%%%%%%%%%%%%%%

%%%%%%%%%%%%%%%%%%%%%%%%%%%%%%%%%%%%%%%%%%%%%%%%%%%%%
\begin{figure}[ht]
\centering
\includegraphics[width=0.48\textwidth]{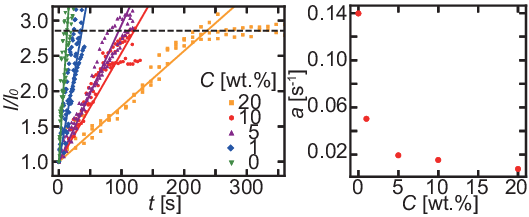}
\caption {\label{swell_speed} (a) Time development of gel particle diameter $l$, where data is normalized by $l_0$, which is the diameter before swelling. The concentration of sodium chloride $C$ is also varied. For each concentration, $C$, three independent measurements are conducted. Initially, the diameter increased linearly. The fitting parameter was obtained from the data between $0 \leq l/l_0 \leq 2.5$, with $l/l_0 = a t+1$. (b) Swelling rate $a$ plotted against $C$.}
\end{figure}
%%%%%%%%%%%%%%%%%%%%%%%%%%%%%%%%%%%%%%%%%%%%%%%%%%%%%

%%%%%%%%%%%%%%%%%%%%%%%%%%%%%%%%%%%%%%%%%%%%%%%%%%%%%
\begin{figure*}[ht]
\centering
\includegraphics[width=0.96\textwidth]{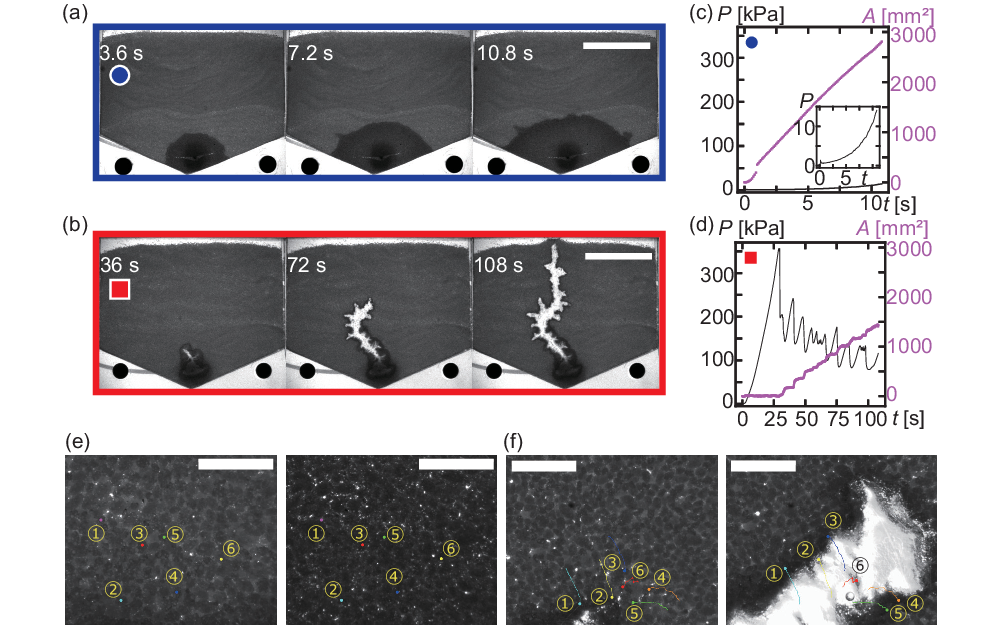}
\caption{\label{timeprop}(a)Snapshots of the dynamics of percolation pattern where $I =$ 10 mL/min. and $C =$ 10 wt.\% (Multimedia view). The injected fluid proceeded in an isotropic manner. (b) Snapshots of clogged pattern where $I =$ 1 mL/min. and $C =$ 0 wt.\% (Multimedia view). The injection front proceeded with branching. The parameters of (a,b) are also indicated in Fig. ~\ref{phase_dia}(a) with a hollow circle and square, respectively. Scale bar: 50 mm. Time development of the infiltrated area $A$ (magenta) and injection pressure $P$ (black) for (c) percolation pattern ($I =$ 10 mL/min. and $C =$ 10 wt.\%) and (d) clogged pattern ($I =$ 1 mL/min. and $C =$ 0 wt.\%).
For the percolation pattern, $P$ and $A$ increases monotonically. On the other hand, for the clogged pattern, $P$ initially increases and then decreases with oscillation. $A$ is increased in a step-wise manner synchronized with the oscillation in $P$. The inset of (c) represents an enlarged plot of $P$ with $t$. Enlarged view of the fluid front where (e) $I =$ 10 mL/min. and $C =$ 10 wt.\%, and (f) $I =$ 0.1 mL/min. and $C =$ 0 wt.\% (Multimedia view). Scale bar: 2 mm. 
Time intervals between the two snapshots are 2 and 120 s, respectively. Lines in the snapshots indicate the trajectories of particles. In the percolation pattern, almost no movement of the particles was observed, whereas the particles were pushed aside by the injection front in the case of the clogged pattern.
}
\end{figure*}
%%%%%%%%%%%%%%%%%%%%%%%%%%%%%%%%%%%%%%%%%%%%%%%%%%%%%
%%%%%%%%%%%%%%%%%%%%%%%%%%%%%%%%%%%%%%%%%%%%%%%%%%%%%
\begin{figure*}[ht]
\centering
\includegraphics[width=0.96\textwidth]{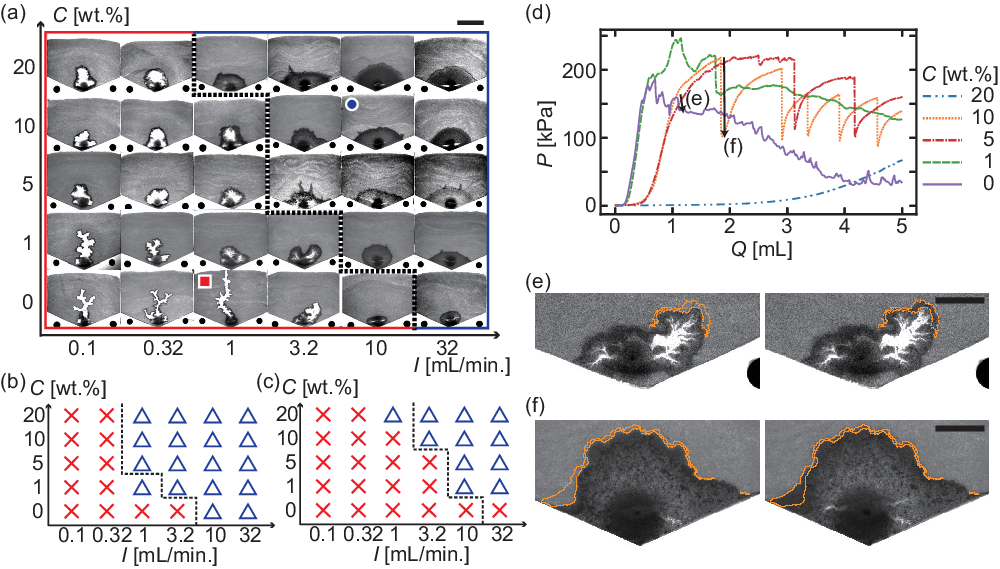}
\caption{\label{phase_dia}Phase diagram of flow pattern. Each snapshot of the flow patterns is for different injection rates $I$ and salinity $C$ of injected fluids. Data were obtained from the identical experiment when the inner fluid is injected with (a) 1.8, (b) 0.9, and (c) 3.6 mL. The dotted line corresponds to the line $\alpha$=1.5. The snapshots in the blue-shaded region (a) and blue triangles (b,c) indicate the percolation pattern, while the snapshots in the red-shaded regions and red crosses indicate the clogged pattern. Hollow circle and square correspond to the parameter in Fig. \ref{timeprop}(a,b). Scale bars correspond to 50 mm. (d) Dependence of pressure on injected volume $Q$ (up to 5 mL) where $C =$0, 1, 5, 10, 20 wt.\% and $I =$3.2 mL/min. We observe a plateau at approximately $P$ = 200 kPa, and a sharp drop in $P$ is noted.
%%%%%%%%%%%%%%%%%%%ここは後で本文に回したほうがいいかも
Snapshots of injection front before and after the pressure drop for $C = $ (e) 0 wt.\% and (f) 10 wt.\%, indicated by arrows in (d) (Multimedia view). Orange lines indicate a swept area in the pressure drops. Scale bars correspond to 20 mm.
}
\end{figure*}
%%%%%%%%%%%%%%%%%%%%%%%%%%%%%%%%%%%%%%%%%%%%%%%%%%%%%

\section{Experimental system}
We used ionic gel particles composed of sodium polyacrylate (AQUA KEEP SA60S, Sumitomo Seika Co., Ltd.) with a medium particle size of 350 $\SI{}{\micro \meter}$. A typical particle is an aggregate of spheres. The sample was stored in a desiccator where the humidity inside was maintained at a low value by desiccant beads of silica gel. Sodium chloride was purchased from FUJIFILM Wako Pure Chemical Cooporation and brilliant blue was purchased from Wako Pure Chemical Industries, Ltd. Pure water, purified using a Millipore MilliQ system, was used.
The injected fluid was an aqueous solution of sodium chloride with concentrations of $C=0, 1, 5, 10$, and $20$ wt.\% and stained with brilliant blue for visualization.
The experimental setup is schematically illustrated in Fig.~\ref{experimentalsystem}.

A Hele-Shaw cell with a gap width of $\SI{1}{\milli \meter}$ was fabricated using acrylic plates (Mitsubishi Rayon Co., Ltd.) with a thickness of $\SI{5}{\milli \meter}$. There was an aperture on the cell through which gases and liquids could exit. An inlet for the injection fluid was constructed with a Luer fitting (ISIS Co., Ltd.) fixed at the upper plate and connected to a nylon tube (Nihon Pisco Co., Ltd.; diameter $\SI{2.5}{\milli \meter}$) with a length of $\SI{570}{\milli \meter}$. A syringe was connected to a nylon tube and installed on a syringe pump (CXF1010; ISIS Co., Ltd.) to inject the inner fluid at a constant injection rate. We used a 50-$\SI{}{\milli \liter}$ syringe (SS-50ESZ; Terumo Co., Ltd.) at an injection rate $I=\SI{32}{\milli \liter / \minute }$. and a 10-$\SI{}{\milli \liter}$ syringe (SS-10SZ; Top Co., Ltd.) at $I$=0.1, 0.32, 1, 3.2, and 10 $\SI{}{\milli \liter / \minute}$. We installed a pressure sensor (26PCDFA6D Honeywell) at the branching point of the nylon tubes made of union T (Nihon, Pisco), 515 mm from the inlet for the conditions where we presented the pressure data. 
The gel particles were then placed at the aperture of the cell. 
To avoid clogging at the injection position, we placed 0.4 mm glass beads (6-257-03, Asone) near the inlet.

As shown in Fig.~\ref{experimentalsystem}(b), 
the fluid was injected using a syringe pump at a fixed injection rate $I$ ($0.1 - 32.0$ $\SI{}{\milli \liter / \minute}$).
The pattern dynamics were recorded from the bottom of the cell using a CMOS camera (DMK 37BUX273; Imaging Source Co.) and analyzed using Image J software \cite{imageJ}. The pressure sensor output was recorded using a data logger (GL7000 with voltage unit GL7-V; GRAPHTEC Co., Ltd.).
Furthermore, the water absorption rate of the gel particles was measured at various NaCl concentrations corresponding to the injected fluid concentrations $C$ as shown in Fig. ~\ref{swell_speed}.
 
Figure~\ref{swell_speed}(a) indicates the normalized particle size $l/l_0$ dependence of time, where $l_0$ is the initial dry particle size. When the particle size reached a cell gap width of $h=$1 $\SI{}{\milli \meter}$, $l/l_0$ reached a dotted line. We fitted the data with a linear fitting function $l/l_0=at+1$ and obtained the swelling rate $a$ as shown in Fig. ~\ref{swell_speed}(b). 
 
\section{Result and Discussion}
Figures~\ref{timeprop}(a) and (b) exemplify the typical behavior of the injected fluid in the granular bed of swelling gel particles (Multimedia view). In the snapshots, light and dark gray and white regions without an apparent texture were observed. Dry gel particles are observed in the light gray region. The injected fluid did not reach this region. The dark gray region represents the fluid and gel particles that were partially swollen by the injected fluid. The white region without an apparent texture represents the fluid and the gel particles that were fully swollen by the injected fluid. 

Snapshots of the percolation patterns observed for high $I$ and $C$ are shown in Fig. ~\ref{timeprop}(a). The front showed small fluctuations, reflecting irregular void spaces; however, the injected fluid flowed in an isotropic manner by passing through the gel particles. The percolation pattern was also characterized by a monotonic increase in the injection pressure $P$ and the infiltrated area $A$, as shown in Fig.~\ref{timeprop}(c). Moreover, $A$ increased steadily up to 3000 mm$^2$, whereas $P$ increased linearly only up to 20 kPa.

Figure~\ref{timeprop}(b) shows snapshots of the clogged pattern, where the swollen gel particle blocks the motion of the fluid and shows the intermittent behavior of the injected fluids, which was observed for low $I$ and low $C$. As shown in the snapshots, we observed anisotropic extension and branching of the injection front.
The clogged pattern is also characterized by an oscillatory drop in the injection pressure after the initial pressure increase, as shown in Fig.~\ref{timeprop}(d). The pressure reached a maximum at $P\sim$ 350 kPa and exhibited a sharp decrease with an amplitude of approximately 100 kPa. The infiltrated area $A$ exhibited a stepwise increase, which was synchronized with the drop in the injection pressure $P$.

To connect the swelling dynamics of each particle to the fluid motion near the injection front, we conducted microscopic observations of the particles, as shown in Fig.~\ref{timeprop}(e) and (f), respectively (Multimedia view). 
Figure~\ref{timeprop}(e) corresponds to the percolation pattern wherein a light gray area turns into a dark gray area, as shown in Fig. ~\ref{timeprop}(a). While the injected fluid percolated through the void space between the gel particles and the light gray region became a dark gray region, the gel particles remained at almost the same position. Care should be taken when gel particles are not fully swollen. Indeed, a fully swollen gel particle does not exhibit a large contrast with the surrounding fluids with respect to the index of refraction. Thus, fully swollen gel particles appeared to be almost invisible in the injected fluid. This is partially confirmed by Fig.~\ref{timeprop}(f), which shows snapshots close to the frontal part of the clogged pattern. The image corresponds to the area near the white-branched region shown in Fig.~\ref{timeprop}(b). In the white region, floating gel particles that were fully swollen by the injected fluids were observed. Moreover, the surrounding partially swollen gel particles blocked the advance of injected fluids. Thus, we found that blocking the fluid flow owing to partially swollen gel particles resulted in the appearance of clogged patterns.

By changing the injection rate $I$ and the salinity of the injected fluid $C$, we constructed phase diagrams of the observed patterns, as shown in Fig.~\ref{phase_dia}. Here, we summarized the data with the same total volume of injected fluids $Q=It$ as $Q=$ (a) 1.8, (b) 0.9, and (c) 3.6 mL. The horizontal and vertical axes correspond to $I$ and $C$, respectively. We obtained a percolation pattern for high $I$ and $C$ and a clogged pattern for low $I$ and $C$. The boundary between the clogged and percolation patterns moved to the upper-right corner (high $I$ and $C$) with increase in $Q$.

Based on the enlarged image shown in Figs.~\ref{timeprop}(e) and (f), the clogged pattern was owing to the blocking of the flow by the partially swollen gel particles. Therefore, we can estimate the phase boundary between the percolation and clogged patterns from the ratio of the characteristic timescales of swelling $t_s$ and injection $t_i$. Here, we considered that $t_s$ from Fig.~\ref{swell_speed} was the time at which the diameter of the gel particles swelled to the gap width of the cell (1 mm). If we consider $t_i$ as the time required for the injected fluid volume to reach 1.8 mL (Fig. ~\ref{phase_dia}(a)), 0.9 mL (Fig. ~\ref{phase_dia}(b)) and 3.6 mL (Fig. ~\ref{phase_dia}(c)), we observed that the line $\alpha$ = $t_s$/$t_i$ = 1.5, which agrees with the phase boundary indicated by the dotted line in Figs. ~\ref{phase_dia}(a)-(c). For $\alpha$ < 1.5, the time is insufficient for the gel particles to block the injected fluid. Consequently, a percolation pattern was observed. With further fluid injection, $t_i$ increased. Then, the region $\alpha$ < 1.5 shifted to the upper-right corner of the phase diagram, as shown by the dotted lines in Fig.~\ref{phase_dia}(a)-(c) indicate. 

An intriguing aspect of the present system is the recursive behavior of the blocking dynamics, which is symbolized by the oscillatory behavior of the injection pressure $P$.The oscillation of $P$ and intermittent propagation of the injection front reflect the finite rigidity of the system.
During the injection process, the swollen particles blocked the fluid flow, the tubes and cells deformed, and the internal pressure increased monotonically. In this process, the injection front supported the increased pressure owing to the friction force between the cell and the swollen particles. 

When the injection pressure reached a critical value that was sufficient to overcome the static friction force acting on the swollen gel particles, the clogging was dissolved by breaking the blocking region, whereas the injected fluid proceeded to the space with dry gel particles. A void space was observed between the dry gel particles; however, eventually, the advanced injected fluids were absorbed by the gel particles to fill the void space. Repeating these processes resulted in the intermittent behavior observed in clogged patterns.

To quantitatively analyze the oscillatory behavior of the clogged patterns, we observed the injection pressure $P$ systematically with $I =$ 3.2 mL/min while changing $C =$ 0, 1, 5, 10, and 20 wt. \%. The results are shown in Fig.~\ref{phase_dia}(d), where $P$ is plotted against the injected volume $Q = It$ (up to 5 mL). %where $C = $0, 1, 5, 10, 20 wt.\% and $I =$ 3.2 mL/min.
At $C$ = 20 wt.\%, the pressure monotonically increased within the observation timescale as typically observed in the percolation pattern. At $C$ = 0, 1, 5, 10 wt.\%, we observed initial pressure increase followed by intermittent pressure drop after the pressure reached the plateau value at approximately $P$ = 200 kPa. This observation corresponded to the typical features of a clogged pattern. The higher the salinity of the injected fluid, the lower the swelling rate, resulting in a first pressure drop at larger $Q$. This result was consistent with the mechanisms discussed above. It should be noted that the percolation pattern should disappear if the injection is continued for an infinite time with infinite-sized flow cells.

In Fig.~\ref{phase_dia}(d), larger pressure drops were observed at high $C$. The difference in these pressure drops reflected the morphology of the clogged pattern. Figures~\ref{phase_dia}(e), (f) show snapshots of the injection front before and after the pressure drop for $C =$ 0 and 10 wt. \%, respectively (Multimedia view). The orange lines indicate the points where the injection front swept during the pressure drop. In both the cases, the pressure drop and fluid propagation coincided. The preceding area and amplitude of the pressure drop for $C =$ 10 wt. \% were larger than those of $C =$ 0 wt.\%. 
This difference occurred because a higher $C$ with a low swelling rate resulted in a larger percolated area before blocking. Thus,
larger areas must slip for a high $C$, and the amplitude of the pressure drop increases.

Additionally, a plateau in the injection pressure at approximately $Q$=2-3 mL at $C =$ 5 and 10 wt. \% was observed. However, the pressure decreased immediately after $P$ reached 200 kPa at $C =$ 0, 1 wt.\%. The plateau in the injection pressure indicates a region with low rigidity. However, our image analysis did not detect the motion of the gel particles before the pressure drop at $C =$ 5 and 10 wt. \%. Thus, the swollen gel particles may deform to form such a small rigidity region; however, microscopic observations may be required.

\section{Summary}
In this study, we observed the flow pattern induced by fluid injection into a Hele-Shaw cell filled with swelling gel particles. The salinity of the injected fluid was varied to control the swelling rate of gel particles to obtain two distinct patterns. For a high salinity level (low swelling rate), high injection rate, and short timescale, the fluid percolated through the granular media and exhibited a percolation pattern with an isotropic front. For a low salinity level (high swelling rate), low injection rate, and long timescale, the fluid pushed aside the swollen particles, which blocked the fluid flow and appeared as a clogged pattern with branching. The injection pressure increased monotonically in the percolation pattern, whereas it exhibited oscillatory behavior in the clogged pattern. Such clogging dynamics were attributed to the repetition between the blocking of the fluid flow and the penetration of the fluid into the blocking region. In addition, the transition from the percolation pattern to the clogged pattern can be understood using the ratio $\alpha$ of the swelling timescale to the injection timescale, which is confirmed by the phase boundary given by $\alpha$ = 1.5. The observation of pressure drops in the clogged pattern indicates that a higher salinity of the injected fluid leads to larger pressure drops at the same injection speed. Furthermore, we elucidated the flow dynamics in media wherein the permeability changed owing to swelling.

Present study exemplified the swelling of granular bed significantly alter the dynamics of injected fluid to give clogged pattern. Interestingly, clogged pattern was characterized with the oscillatory repetitive behavior in pressure drop and appearance of blocking region. Previous studies in geoscience suggest that the oscillatory behavior of the pore fluid pressure can reduce the seismic efficiency of the injection-induced earthquakes~\cite{Zang2019}. Thus, autonomous appearance of the oscillatory pressure change in the clogged pattern may have impact on the injection-induced earthquake. Furthermore, the experimental results showing that the larger amplitude of pressure oscillations appeared for lower swelling rate may have a counter-intuitive but relevant insight for such geological situations. 

In this study, we have investigated experiments with macroscopic pattern formation owing to the injected fluid. We also gave the phenomenological explanation about mechanism for the clogging dynamics. To go beyond such a phenomenological argument, we may conduct the fully quantitative study on the rheological behavior of granular particles with active swelling dynamics. For example, the dynamic rheological measurements of swollen particles including their swelling dynamics should be of great interest. Numerical simulation based on the microscopic dynamics of swelling granular particles would be also helpful.

%%%%%%%%%%%%%%%%%%%%%%%%%%%%%%
\section*{ACKNOWLEDGMENTS}
This study was partially supported by a Grant-in-Aid from the Japan Society for the Promotion of Science, KAKENHI (Grant No. \ JP16H06478, No.\ JP21H01004 and No.\ JP21H01006), JSPS Research Fellow (Grant No.\ JP21J13720),  Leave a Nest Co., ltd., JST, the establishment of university fellowships towards the creation of science technology innovation (Grant No. JPMJFS2144) and the JSPS Core-to-Core Program “Advanced core-to-core network for the physics of self-organizing active matter (JPJSCCA20230002)”. We would like to thank Editage (www.editage.jp) for English language editing.
%%%%%%%%%%%%%%%%%%%%%%%%%%%%%%


\end{document}